\documentclass[onecollarge,fleqn,runningheads, preprint,showpacs,preprintnumbers,amsmath,amssymb]{svjour2}
\smartqed  
\usepackage{graphicx} 
\usepackage{dcolumn} 
\usepackage{bm} 
\begin{document}

\title{Refinement of a previous hypothesis of the Lyapunov analysis of isotropic turbulence
}


\author{Nicola de Divitiis
}


\institute{Dipartimento di Ingegneria Meccanica e Aerospaziale\\
"La Sapienza" University, Rome, Italy \at
           via Eudossiana, 18  \\
              Tel.: +39-06-44585268\\
              Fax: +39-06-4881759\\
              \email{n.dedivitiis@gmail.com}           
}

\date{Received: date / Accepted: date}

\maketitle

\newcommand{\no}{\noindent}
\newcommand{\be}{\begin{equation}}
\newcommand{\ee}{\end{equation}}
\newcommand{\bea}{\begin{eqnarray}}
\newcommand{\eea}{\end{eqnarray}}
\newcommand{\bc}{\begin{center}}
\newcommand{\ec}{\end{center}}

\newcommand{\calr}{{\cal R}}
\newcommand{\calv}{{\cal V}}

\newcommand{\bff}{\mbox{\boldmath $f$}}
\newcommand{\bfg}{\mbox{\boldmath $g$}}
\newcommand{\bfh}{\mbox{\boldmath $h$}}
\newcommand{\bfi}{\mbox{\boldmath $i$}}
\newcommand{\bfm}{\mbox{\boldmath $m$}}
\newcommand{\bfp}{\mbox{\boldmath $p$}}
\newcommand{\bfr}{\mbox{\boldmath $r$}}
\newcommand{\bfu}{\mbox{\boldmath $u$}}
\newcommand{\bfv}{\mbox{\boldmath $v$}}
\newcommand{\bfx}{\mbox{\boldmath $x$}}
\newcommand{\bfy}{\mbox{\boldmath $y$}}
\newcommand{\bfw}{\mbox{\boldmath $w$}}
\newcommand{\bfk}{\mbox{\boldmath $\kappa$}}

\newcommand{\bfA}{\mbox{\boldmath $A$}}
\newcommand{\bfD}{\mbox{\boldmath $D$}}
\newcommand{\bfI}{\mbox{\boldmath $I$}}
\newcommand{\bfL}{\mbox{\boldmath $L$}}
\newcommand{\bfM}{\mbox{\boldmath $M$}}
\newcommand{\bfS}{\mbox{\boldmath $S$}}
\newcommand{\bfT}{\mbox{\boldmath $T$}}
\newcommand{\bfU}{\mbox{\boldmath $U$}}
\newcommand{\bfX}{\mbox{\boldmath $X$}}
\newcommand{\bfY}{\mbox{\boldmath $Y$}}
\newcommand{\bfK}{\mbox{\boldmath $K$}}

\newcommand{\bfrho}{\mbox{\boldmath $\rho$}}
\newcommand{\bfchi}{\mbox{\boldmath $\chi$}}
\newcommand{\bfphi}{\mbox{\boldmath $\phi$}}
\newcommand{\bfPhi}{\mbox{\boldmath $\Phi$}}
\newcommand{\bflambda}{\mbox{\boldmath $\lambda$}}
\newcommand{\bfxi}{\mbox{\boldmath $\xi$}}
\newcommand{\bfLambda}{\mbox{\boldmath $\Lambda$}}
\newcommand{\bfPsi}{\mbox{\boldmath $\Psi$}}
\newcommand{\bfomega}{\mbox{\boldmath $\omega$}}
\newcommand{\bfeps}{\mbox{\boldmath $\varepsilon$}}
\newcommand{\bfepsn}{\mbox{\boldmath $\epsilon$}}
\newcommand{\bfzeta}{\mbox{\boldmath $\zeta$}}
\newcommand{\bfkappa}{\mbox{\boldmath $\kappa$}}
\newcommand{\bfsigma}{\mbox{\boldmath $\sigma$}}
\newcommand{\itPsi}{\mbox{\it $\Psi$}}
\newcommand{\itPhi}{\mbox{\it $\Phi$}}
\newcommand{\bint}{\mbox{ \int{a}{b}} }
\newcommand{\ds}{\displaystyle}
\newcommand{\Sum}{\Large \sum}

\begin{abstract}
The purpose of this brief comunication is to improve a hypothesis of the previous work of the author (de Divitiis, Theor Comput Fluid Dyn, doi:10.1007/s00162-010-0211-9) dealing with the finite--scale Lyapunov analysis of isotropic turbulence. There, the analytical expression of the structure function of the longitudinal velocity difference $\Delta u_r$ is derived through a statistical analysis of the Fourier transformed Navier-Stokes equations, and 
by means of considerations regarding the scales of the velocity fluctuations, which arise from the Kolmogorov theory. Due to these latter considerations, this Lyapunov analysis seems to need some of the results of the Kolmogorov theory.

This work proposes a more rigorous demonstration which leads to the same 
structure function, without using the Kolmogorov scale. 
This proof assumes that pair and triple longitudinal correlations are sufficient to
determine the statistics of $\Delta u_r$, and adopts a reasonable canonical decomposition of the velocity difference in terms of proper stochastic variables which are adequate to describe the mechanism of kinetic energy cascade.
\end{abstract}

\section{\bf Introduction}

In the previous work \cite{deDivitiis_1}, the author applies the finite--scale Lyapunov theory to analyse the homogeneous isotropic turbulence.
In particular, this theory leads to the analytical closure of
the von K\'arm\'an--Howarth equation, giving the longitudinal triple velocity correlation $k$ in terms of the longitudinal velocity correlation $f$ and $\partial f /\partial r$ 
(see also the Appendix), and shows that the structure function of the longitudinal velocity difference is expressed by  
\bea
\begin{array}{l@{\hspace{+0.2cm}}l}
\ds \frac{\Delta u_r}  {\sqrt{\langle (\Delta u_r)^2} \rangle} =
\ds \frac{   {\xi} + \psi \left( \chi ( {\eta}^2-1 )  -  
\ds  ( {\zeta}^2-1 )  \right) }
{\sqrt{1+2  \psi^2 \left( 1+ \chi^2 \right)} } 
\end{array}
\label{fluc intro}
\eea 
where $\Delta u_r = ({\bf u}({\bf x}+ {\bf r}) - {\bf u}({\bf x})) \cdot {\bf r}/r$, $\xi$, $\eta$ and $\zeta$ are uncorrelated centered gaussian random variables, with 
$\langle \xi^2 \rangle = \langle \eta^2 \rangle = \langle \zeta^2 \rangle$ = 1,  and $\psi$ is a function of the Taylor scale Reynolds number $R_\lambda = {u \lambda_T}/{\nu}$ and of the separation distance $r \equiv \vert {\bf r} \vert$, according to
\bea
\psi(r, R_\lambda) =  
\sqrt{\frac{R_\lambda}{15 \sqrt{15}}} \ \hat{\psi}(r),
\label{Rl intro}
\eea
being $u=\sqrt{\langle u_r^2 \rangle}$ the longitudinal velocity standard deviation,
$\lambda_T$ is the Taylor microscale,
and $\chi = \chi (R_\lambda)$ $\ne 1$ is a proper function of $R_\lambda$ which provides nonzero skewness of $\Delta u_r$ \cite{deDivitiis_1}, \cite{deDivitiis_2}.

In Ref. \cite{deDivitiis_1}, the demonstration of Eq. (\ref{fluc intro})
is carried out through statistical elements regarding the Fourier transformed Navier--Stokes equations, whereas the proof of Eq. (\ref{Rl intro}) is based on the fact that, according to the Kolmogorov theory, the ratio (small scale velocity)--(large scale velocity) depends on 
$\lambda_T/\ell \approx \sqrt{R_\lambda}$, where $\ell$ is the Kolmogorov microscale.
Therefore, the analysis of Ref. \cite{deDivitiis_1} seems to require 
the adoption of $\ell$ whose definition is based on another theory.

\bigskip

Here, instead of using the Kolmogorov scale, we obtain Eqs. (\ref{fluc intro}) and (\ref{Rl intro}), starting from the canonical decomposition of the fluid velocity in terms of proper centered random variables $\xi_k$. 
In order to describe the mechanism of kinetic energy cascade, the variables $\xi_k$ are properly chosen in such a way that each of them exhibits a non--symmetric distribution function.
Moreover, due to the isotropy, we postulate that the knowledge of $f$ and $k$ represents a 
sufficient condition to determine the statistics of $\Delta u_r$.

\bigskip

\section{\bf Lyapunov analysis of the velocity fluctuations}

This section renews the procedure for calculating the velocity fluctuations, 
which is based on the Lyapunov analysis of the fluid strain \cite{deDivitiis_1},
and on the momentum Navier-Stokes equations
\bea
\ds \frac{\partial  {u}_k}{\partial t}  = 
- \frac{\partial  {u}_k}{\partial x_h} u_h +
\frac{1}{\rho}  \frac{\partial T_{k h}}  {\partial x_{h}} 
\label{N-S}
\label{P}
\eea
being ${\bf u} \equiv (u_1, u_2, u_3)$, $T_{k h}$ and $\rho$ the fluid velocity,  
stress tensor and density, respectively.

In order to obtain the velocity fluctuation, consider now the relative motion between two contiguous particles, expressed by the infinitesimal separation vector $d {\bf x}$ which obeys to the equation
\bea
d \dot{\bf x} = \nabla {\bf u} \ d {\bf x}
\label{dx}
\eea 
where $d{\bf x}$ varies according to the velocity gradient which in turn follows the Navier-Stokes equations.
As observed in Ref. \cite{deDivitiis_1}, $d{\bf x}$ is much faster than the fluid state variables, and the Lyapunov analysis of Eq. (\ref{dx}) provides the expression of the local deformation in terms of maximal Lyapunov exponent $\Lambda >0$
\bea
\frac{\partial {\bf x}}{\partial {\bf x}_0} \approx {\mbox e}^{\Lambda (t - t_0)} 
\label{stretch}
\eea
The map $\bfchi$ : ${\bf x}_0 \rightarrow {\bf x}$, 
is the function which gives the current position $\bf x$ of a fluid particle located at the referential position ${\bf x}_0$ at $t = t_0$ \cite{Truesdell77}.
Equation (\ref{P}) can be written in terms of the referential position 
${\bf x}_0$ \cite{Truesdell77} 
\bea
\ds \frac{\partial  {u}_k}{\partial t}  =  \left( -\frac{\partial  {u}_k}{\partial x_{0 p}} u_h +
\frac{1}{\rho}
 \frac{\partial T_{k h}}  {\partial x_{0 p}} \right)  \ \frac{\partial x_{0 p}}{\partial x_{h}} 
\label{N-Sr}
\eea
The adoption of the referential coordinates allows to factorize the velocity fluctuation and to express it in Lyapunov exponential form of the local deformation.
 As this deformation is assumed to be much more rapid than 
$ -{\partial  {u}_k}/{\partial x_{0 p}} u_h +
1/\rho \  {\partial T_{k h}} / {\partial x_{0 p}}$,
the velocity fluctuation can be obtained integrating Eq. (\ref{N-Sr}) 
with respect to the time, where 
$ -{\partial  {u}_k}/{\partial x_{0 p}} u_h + 1/\rho \ {\partial T_{k h}}/{\partial x_{0 p}}$ is considered to be constant 
\bea
\begin{array}{l@{\hspace{0cm}}l}
\ds u_k \approx 
\frac{1}{\Lambda}  \ 
\left( -\frac{\partial  {u}_k}{\partial x_{0 p}} u_h +
\frac{1}{\rho}
 \frac{\partial T_{k h}}  {\partial x_{0 p}} \right)_{t = t_0} 
\end{array}
\label{fluc_v2_0}
\label{fluc_v2}
\eea
This assumption is justified by the fact that, according to the classical 
formulation of motion of continuum media \cite{Truesdell77}, the terms into the circular brackets of Eq. (\ref{N-Sr}) are considered to be smooth functions of $t$ -at least during the period of a fluctuation-  whereas the fluid deformation varies very rapidly according to Eqs. (\ref{dx})-(\ref{stretch}).

\bigskip

\section{\bf Statistical analysis of velocity difference}

As explained in this section, the Lyapunov analysis of the local deformation
and some plausible assumptions about the statistics of $\bf u$ 
lead to determine the structure function of $\Delta u_r$ and its PDF.

The statistical properties of $\Delta u_r$ are here investigated expressing the 
fluid velocity  through the following canonical decomposition \cite{Ventsel}
\bea
{\bf u} = \Sum_k  \hat{\bf U}_k \xi_k, 
\label{X0}
\eea 
where $\hat{\bf U}_k$ ($k=1, 2, ...$) are proper coordinate functions of $t$ and $\bf x$, and $\xi_k$ ($k=1, 2, ...$) are certain dimensionless independent stochastic variables which satisfy
\bea
\left\langle \xi_k \right\rangle = 0, \ \ \ 
\left\langle \xi_i \xi_j \right\rangle = \delta_{i j}, \ \ \ 
\left\langle \xi_i \xi_j \xi_k  \right\rangle =  \varpi_{i j k} \ p, \ \ \ 
 \vert p \vert >>> 1, \ \
\vert p \vert >>> \left\langle \xi_k^4 \right\rangle
\label{X1}
\eea
where $\varpi_{i j k}$ = 1 for $i = j = k$, else $\varpi_{i j k}$=0.
It is worth to remark that the variables $\xi_k$ are adequately chosen in such a way that they can describe properly the mechanism of energy cascade. Specifically, the adoption of $\xi_k$ with $\vert p \vert > > >$ 1 is justified by the fact that the evolution equation of the velocity correlation (see for instance the von K\'arm\'an-Howarth equation, appendix) includes also the third order velocity correlation
$k(r)$ which is responsible for the intensive mechanism of energy cascade and 
$\langle (\Delta u_r)^3 \rangle/ \langle (\Delta u_r)^2 \rangle^{3/2} \ne$ 0. 
As the result, it is reasonable that the canonical decomposition (\ref{X0}),
($\Delta u_r = ({\bf r}/r) \cdot \Sum_k (\hat{\bf U}_k({\bf x} + {\bf r})- \hat{\bf U}_k({\bf x} ))  \xi_k$) includes variables $\xi_k$ with $\vert \langle \xi_k^3 \rangle \vert >>>$ 1 \cite{Lehmann99}.
This has very important implications for what concerns the statistics of the fluctiations
of $\Delta u_r$.
In order to analyze this question, consider now the dimensionless velocity fluctuation $\hat{\bf u}$. This is obtained in terms of $\xi_k$ substituting Eq. (\ref{X0}) into Eq. (\ref{fluc_v2})
\bea
\hat{u}_h =  \Sum_{i j} A_{h i j} \xi_i \xi_j 
+ \frac{1}{R_\lambda} \Sum_{k} b_{h k} \xi_k  
\label{delta u}
\eea
where $r = \hat{r} \lambda_T$ and $u_h = \hat{u}_h  u$.
Therefore $\Sum_{i j} A_{h i j} \xi_i \xi_j$ arises from the 
inertia and pressure terms, whereas  $1/R_\lambda \Sum_{k} b_{h k} \xi_k$ is due 
to the fluid viscosity.
Now, thanks to the local isotropy, $u_h$ is a gaussian stochastic variable \cite{Lehmann99}, \cite{Ventsel},  accordingly, $\xi_k$ satisfy into Eq. (\ref{delta u}), the Lindeberg condition, a very general necessary and sufficient condition for satisfying the central limit theorem \cite{Lehmann99}. 
This condition does not apply to the velocity difference.
In fact, as $\Delta {\bf u}$ is the difference between two correlated gaussian variables, its PDF could be a non gaussian distribution function.
To study this, the fluctuation $\ds \Delta u_r$ is first expressed in terms of $\xi_k$
\bea
\Delta \hat{u}_r ({\bf r}) 
 =   \Sum_{i j} \Delta A_{r i j}({\bf r}) \xi_i \xi_j 
+ \frac{1}{R_\lambda}  \Sum_{k} \Delta b_{r k} ({\bf r}) \xi_k 
   \equiv L + S + G^+ + G^-
\label{fluc T}
\eea
This fluctuation can be reduced to the contributions $L$,  $S$, $G^+$ and $G^-$, appearing into Eq. (\ref{fluc T}) \cite{Madow40}:
in particular, $L$ is the sum of all linear terms due to the fluid viscosity, 
$S \equiv S_{i j} \xi_i \xi_j$ is the sum of all bilinear forms arising from the inertia and pressure terms, whereas $G^+$ and $G^-$ are, respectively, definite positive and negative quadratic forms of centered gaussian variables, which derive from the inertia and pressure terms.
The quantity $L+S$ tends to a gaussian random variable being the sum of statistically orthogonal terms \cite{Madow40}, \cite{Lehmann99}, while $G^+$ and $G^-$ are determined by means of the hypotheses of isotropy and of fully developed flow
\bea
\begin{array}{l@{\hspace{+0.2cm}}l}
 G^- = -(\zeta^2 -1) \psi_2(r), \\\\
 G^+ = (\eta^2 -1) \psi_3(r)   
\end{array} 
\label{nn}
\eea
Observe that, due to these hypotheses, $G^+$ and $G^-$ are uncorrelated, 
thus $\eta$, $\zeta$ are two independent centered gaussian variables, with 
$\langle \eta^2 \rangle = \langle \zeta^2 \rangle$ =1.
Furthermore, as the knowledge of $f$ and $k$ is considered to be a sufficient
condition for determining the statistics of $\Delta u_r$, $\psi_2$ and $\psi_3$ are assumed to be proportional with each other through a constant which depends only on $R_\lambda$
\bea
\chi(R_\lambda) = \frac{\psi_3(r)}{\psi_2(r)}
\eea
Therefore, the longitudinal velocity difference can be written as
\bea
\begin{array}{l@{\hspace{+0.2cm}}l}
\ds \Delta u_r  = 
\psi_1 {\xi} + 
\ds \psi_2 \left( \chi  ( {\eta}^2-1 )  - ( {\zeta}^2-1 )  \right) 
\end{array}
\label{fluc3}
\eea
where $\xi$ is a centered gaussian random variable with 
$\langle \xi^2 \rangle$ = 1, that thanks to the hypotheses of fully developed flow
and of isotropy, is considered to be statistically independent from $\eta$ and $\zeta$.

Comparing the terms of Eqs. (\ref{fluc3}) and (\ref{fluc T}),
we obtain that $\psi_1$ and $\psi_2$ are related with each other and that their ratio 
$\psi \equiv \psi_1/ \psi_2$ depends on $R_\lambda$ and $r$
\bea
2 (1+ \chi^2) \frac{\psi_2^2 }{ \psi_1^2} =
\frac{ \left\langle (G^+ + G^-)^2 \right\rangle}{\left\langle ( S_{i j} \xi_i \xi_j 
+ 1/R_\lambda  \Delta b_{r k} \xi_k )^2 \right\rangle}
\label{psi_ratio}
\eea
Now, the divisor at the R. H. S. of Eq. (\ref{psi_ratio}) is the sum
of the following three terms:
\bea
\ds A = S_{i j} S_{p q} \langle \xi_i \xi_j \xi_p \xi_q \rangle, \ \ 
\ds B= \frac{2}{R_\lambda} S_{i j} \langle \xi_i \xi_j \xi_k \rangle 
\Delta b_{r k},  \ \ 
\ds C= \frac{1}{R_\lambda^2} \Delta b_{r k}^2 \langle \xi_k^2 \rangle
\eea
Hence, taking into account the properties (\ref{X1}) of $\xi_k$, 
$\vert B \vert >>>$ $\vert A \vert$, $\vert C \vert$, thus
$\psi$ tends to a quantity arising only from the terms $\langle \xi_k^3 \rangle$
which appear in Eq. (\ref{psi_ratio}).
\bea
\psi \equiv \frac{\psi_2}{ \psi_1} = \varphi(r) \sqrt{R_\lambda}
\label{Rl 0}
\eea 
This expression corresponds to that obtained in Ref. \cite{deDivitiis_1}
\bea
\psi(r, R_\lambda) =  
\sqrt{\frac{R_\lambda}{15 \sqrt{15}}} \
\hat{\psi}(r), \ \ \ \ \hat{\psi}(0) = O(1)
\label{Rl bis}
\eea
and the dimensionless longitudinal velocity difference is given by Eq. (\ref{fluc intro}) 
\bea
\begin{array}{l@{\hspace{+0.2cm}}l}
\ds \frac {\Delta u_r}{\sqrt{\langle (\Delta u_r)^2} \rangle} =
\ds \frac{   {\xi} + \psi \left( \chi ( {\eta}^2-1 )  -  
\ds  ( {\zeta}^2-1 )  \right) }
{\sqrt{1+2  \psi^2 \left( 1+ \chi^2 \right)} } 
\end{array}
\label{fluc4 bis}
\eea
It is worth to remark that $\psi$ expresses the fluctuations ratio 
(large scale velocity)--(small scale velocity), that is  
$\psi \approx u/u_s \approx (u^2/\lambda_T) (l_s/u_s^2)$, where $l_s$ and $u_s$ are, respectively the characteristic small scale and the corresponding velocity.
This implies that $u/u_s \simeq \lambda_T/l_s \approx \sqrt{R_\lambda}$, thus
$l_s$ identifies the Kolmogorov scale and $u_s l_s/\nu \approx 1$ is the
corresponding Reynolds number.

The distribution function of $\Delta u_r$ is then 
expressed through the Frobenius-Perron equation \cite{Nicolis95}, taking into account that 
$\xi$, $\eta$ and $\zeta$ are independent stochastic variables
\bea
\begin{array}{l@{\hspace{+0.0cm}}l}
F(\Delta u_r') = \hspace{-0.mm}
\ds \int_\xi \hspace{-0.mm}
\int_\eta  \hspace{-0.mm}
\int_\zeta \hspace{-0.mm}
p(\xi) p(\eta) p(\zeta) \
\delta \left( \Delta u_r'\hspace{-0.mm}-\hspace{-0.mm} \Delta u_r(\xi, \eta ,\zeta) \right)
d \xi \ d \eta \ d \zeta
\end{array}
\label{Tfrobenious_perron}
\eea 
where $\Delta u_r(\xi, \eta ,\zeta)$ is determined with Eq. (\ref{fluc4 bis}), $\delta$ is the Dirac delta and $p$ is a centered gaussian PDF with standard deviation equal to the unity.
The dimensionless statistical moments of $\Delta u_r$ are easily calculated
\bea
\begin{array}{l@{\hspace{+0.2cm}}l}
\ds H_n \equiv \frac{\left\langle (\Delta u_r )^n \right\rangle}
{\left\langle (\Delta u_r)^2 \right\rangle^{n/2} }
= 
\ds \frac{1} {(1+2(1+\chi^2)  \psi^2)^{n/2}} 
\ds \sum_{k=0}^n 
\left(\begin{array}{c}
n  \\
k
\end{array}\right)  \psi^k
 \langle \xi^{n-k} \rangle 
  \langle (\chi (\eta^2-1)  - (\zeta^2-1) )^k \rangle 
\end{array}
\label{Tm1}
\eea
In particular, $H_3$, related to the mechanism of energy cascade, is
\bea
\ds H_3(r) = \frac{8 \psi^3(\chi^3-1)}{(1+ 2 \psi^2 (1+\chi^2))^{3/2}} 
\eea

In conclusion, $\chi = \chi(R_\lambda)$ is implicitly calculated, in function of  $\psi(0)$, 
taking into account that this Lyapunov theory gives $H_3(0) = -3/7$ (see Appendix)
\cite{deDivitiis_1}
\bea
\ds H_3(0) = \frac{8 \psi^3(0)(\chi^3-1)}{(1+ 2 \psi^2(0) (1+\chi^2))^{3/2}} = - \frac{3}{7}
\eea
where $\hat{\psi}(0) \simeq 1.075$ is estimated as in Ref. \cite{deDivitiis_1},
and  $\sqrt{\langle (\Delta u_r)^2 \rangle}$ and $\hat{\psi}(r)$ are calculated
in function of $f(r)$ and $k(r)$.

\bigskip

We conclude this brief comunication by observing that the mechanism of energy cascade
acts on $\Delta u_r$ whose expression, here calculated with the finite-scale Lyapunov theory and Eq. (\ref{X1}), provides a non-symmetric PDF, where the absolute values of the dimensionless moments $\vert H_n(0) \vert$, rise with the Taylor scale Reynolds number for $n >$ 3.

\bigskip

\section{\bf  Acknowledgments}

This work was partially supported by the Italian Ministry for the 
Universities and Scientific and Technological Research (MIUR).

\bigskip

\section{\bf Appendix}

For sake of convenience, this section reports some of the results dealing with the closure of the von K\'arm\'an-Howarth equation, obtained in Refs. \cite{deDivitiis_1} and  \cite{deDivitiis_2}.

For fully developed isotropic homogeneous turbulence, the pair correlation
function 
\bea
f(r) = \frac{\langle u_r({\bf x}) u_r({\bf x}+ {\bf r}) \rangle}{u^2}
\eea
satisfies the von K\'arm\'an-Howarth equation \cite{Karman38}
\bea
\ds \frac{\partial f}{\partial t} = 
\ds  \frac{K(r)}{u^2} +
\ds 2 \nu  \left(  \frac{\partial^2 f} {\partial r^2} +
\ds \frac{4}{r} \frac{\partial f}{\partial r}  \right) +\frac{10 \nu}{\lambda_T^2} f 
\label{vk-h}
\eea
the boundary conditions of which are
\bea
\begin{array}{l@{\hspace{+0.2cm}}l}
\ds f(0) = 1,  \\\\
\ds \lim_{r \rightarrow \infty} f (r) = 0
\end{array}
\label{bc0}
\eea
where $u \equiv \sqrt{\langle u_r^2({\bf x}) \rangle}$ 
satisfies the equation of the turbulent kinetic energy \cite{Karman38}
\bea
\ds \frac{d u^2}{d t} = - \frac{10 \nu}{\lambda_T^2} u^2 
\eea
and $\lambda_T \equiv \sqrt{-1/f''(0)}$ is the Taylor scale.
The function $K(r)$, giving  the mechanism of energy cascade, is related to the 
longitudinal triple velocity correlation function $k$ 
\bea
\begin{array}{l@{\hspace{+0.0cm}}l}
\ds K(r) = u^3 \left( \frac{\partial }{\partial r} + \frac{4}{r} \right) k(r), 
\ \ \mbox{where} \ \ 
\ds k(r) = \frac{\langle u_r^2({\bf x}) u_r({\bf x}+ {\bf r}) \rangle}{u^3}
\end{array}
\eea
Thus, the von K\'arm\'an-Howarth equation provides the relationship between the statistical
moments $\left\langle (\Delta u_r)^2 \right\rangle$ and $\left\langle (\Delta u_r)^3 \right\rangle$ in function of $r$.

The Lyapunov theory proposed in Ref. \cite{deDivitiis_1} gives the
closure of the von K\'arm\'an-Howarth equation, and expresses  
$K(r)$ in terms of $f$ and $\partial f/\partial r$ 
\bea
\ds K(r) = u^3 \sqrt{\frac{1-f}{2}} \frac{\partial f}{\partial r}
\label{K}
\label{K closure}
\eea
Accordingly, the skewness of$\Delta u_r$ is \cite{Batchelor53}
\bea
\ds H_3(r) \equiv 
\frac{\langle (\Delta u_r)^3 \rangle }{\langle (\Delta u_r)^2 \rangle^{3/2}} 
=
\frac{6 k(r)}{(2(1-f(r)))^{3/2}}
\label{H3}
\eea
Therefore, the skewness of $\partial u_r/ \partial r$ is 
\bea
H_3(0) = -\frac{3}{7}
\eea

\bigskip




\end{document}